\documentclass[letters, 10pt]{fcs}

\usepackage{mathtools}
\usepackage{subfigure}
\def\model{scInterpreter}

\title{scInterpreter: Training Large Language Models to Interpret scRNA-seq Data for Cell Type Annotation}
\author[2,3]{Cong Li}
\author*[1]{Meng Xiao}
\author[1,3]{Pengfei Wang}
\author[2]{Guihai Feng}
\author[2]{Xin Li}
\author[1,3]{Yuanchun Zhou}
\address[1]{Computer Network Information Center, Chinese Academy of Sciences, Beijing,
100083, China}
\address[2]{State Key Laboratory of Stem Cell and Reproductive Biology, Institute of Zoology, Chinese Academy of Sciences, Beijing, 100101, China}
\address[3]{University of Chinese Academy of Sciences, Beijing, 100864, China}
\fcssetup{
  received       = {month dd, yyyy},
  accepted       = {month dd, yyyy},
  corr-email     = {shaow@cnic.cn},
}
\begin{abstract}
Despite the inherent limitations of existing Large Language Models in directly reading and interpreting single-cell omics data, they demonstrate significant potential and flexibility as the Foundation Model. This research focuses on how to train and adapt the Large Language Model with the capability to interpret and distinguish cell types in single-cell RNA sequencing data. Our preliminary research results indicate that these foundational models excel in accurately categorizing known cell types, demonstrating the potential of the Large Language Models as effective tools for uncovering new biological insights.
\end{abstract}

\begin{document}

\section{Introduction}

Recent studies~\cite{cui2023scgpt,hao2023scfoundation,theodoris2023geneformer,yang2023genecompass} make significant progress in learning gene-level and cell-level representation by leveraging large-scale gene expression data. 

However, there are few studies~\cite{chen2023genept} that focus on adopting the external description and common knowledge from Large Language Models 
 (LLMs) to enhance the representation. 
Despite the inherent limitations of existing LLMs~\cite{ye2023character}  in directly reading and interpreting gene expression data, they demonstrate significant potential and flexibility as the Foundation Model. 
This research proposes \textbf{\model}, which focuses on utilizing the LLMs with the capability to directly interpret and distinguish cell types in gene expression data. 
Our preliminary experiments indicate that \model excels in accurately categorizing known cell types, demonstrating the potential of the common knowledge in LLMs as an effective bridge for uncovering biological insights.

\begin{figure*}
\centering
\includegraphics[width=\textwidth]{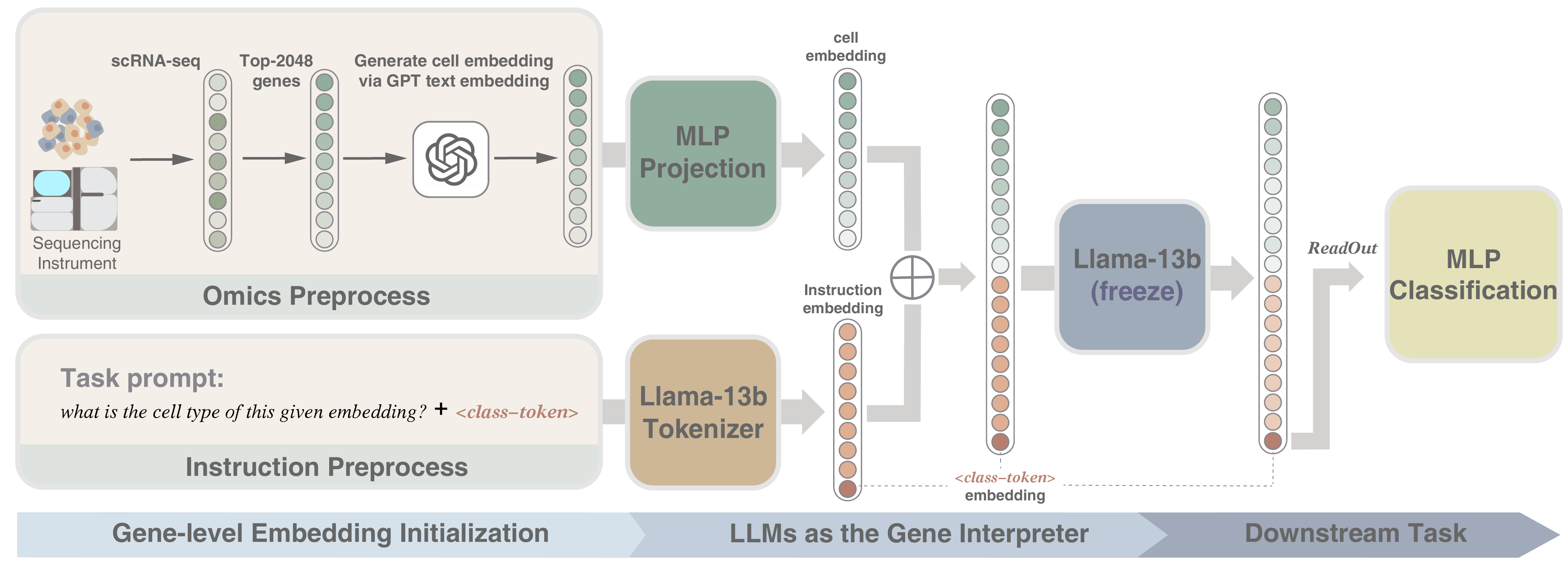}
\caption{The pipeline of scInterpreter. The model will first embed each input from the cell and downstream task-specific instruction. Then the cell embedding and instruction embedding will pass through the LLMs. After aggregating the knowledge and structural information of the given cell by LLMs, the model \textit{ReadOut} the representation and then conducts the downstream task.}
\vspace{-0.3cm}
\label{main}
\end{figure*}

\section{Methodology}
Our proposed method consists of two main parts. 
The first part aims to generate the gene-level representation via LLMs based on the descriptive text of each gene.
The target of the second part is to generate the cell embedding via the specific gene expression and LLM.
After these stages, we could obtain the cell embedding and then feed it into the downstream task, such as cell-type annotation. 
\subsection{Gene-level Embedding Initialization}
The representation of each gene is initialized with its descriptive texts, which is extracted from the NCBI dataset\footnote{\url{https://www.ncbi.nlm.nih.gov/gene}}.
Those texts will then be fed into the GPT-3.5\footnote{\url{https://api.openai.com/v1/embeddings}} to generate the embeddings for each gene, defined as:
\begin{equation}
    e = f_{gpt}(T_{des}),
\end{equation}
where $T_{des}$ is the description of a given gene, $f_{gpt}(\cdot)$ represents the \textit{text-embedding-ada-002} model in \\GPT-3.5, and $e$ is the representation of this gene. 


\subsection{LLMs as the Gene Interpreter}
We select Llama-13b~\cite{touvron2023llama} as the based LLM. 
The cell representation then projects to 5120 dimensions through a multi-layer perceptron (MLP) containing to conform to the Llama-13b's input dimensions $h$:
\begin{equation}
    C = MLP_p(e_1\oplus e_2\oplus \cdots \oplus e_n),
\end{equation}
where $e_1$ to $e_n$ represents the ranked top-$n$ expressed genes' generated embeddings in this cell. $\oplus$ is the element-wise concatenate. $MLP_p$ is the projection layer with the learnable parameter. $C \in \mathbf{R}^{n\times h}$ is the embedding of the given cell. 

We then pass the cell embedding matrix into the Llama-13b along with the downstream task instruction, such as \textit{`what is the cell type of this given embedding?'}. After that, we take the \textit{class-token} from the output and feed it into a trainable classification head:
\begin{align}
    E_{ins} \oplus C \oplus e_{cls} \xrightarrow[ReadOut]{LLM} \hat{e}_{cls}, \\
    \hat{y} = \text{Softmax}(MLP_{c}(\hat{e}_{cls})),
\end{align}
where $E_{ins}$ is the text embedding of the given instruction, $e_{cls}$ is the embedding of the class-token. After fed into LLM, scInterpreter will $ReadOut$ the output. For the cell-type annotation task, we set the $ReadOut$ operation as directly taking the class-token embedding from the output. 
$\hat{e}_{cls}$ is the class-token embedding after the aggregation within the LLM. $\hat{y}$ is the model prediction.
During the training process, the Llama model will be frozen. The projection layer, the classification head, and the \textit{class-token} token's embedding layer will be optimized by the cross-entropy loss.

\section{Experiment}
\subsection{Datasets Description}
We construct two scRNA-seq datasets. 
HUMAN-10k comprises 10,000 single-cell sequencing records with 61 different cell types, each having 23,111 genes gene representations. 
MOUSE-13k comprises 13,000  records with 37 different cell types, each having 27,443 gene representations.

\subsection{Study of the Cell-type Annotation}
\begin{figure}
\centering
\subfigure[HUMAN-10k]{
    \includegraphics[width=0.45\linewidth]{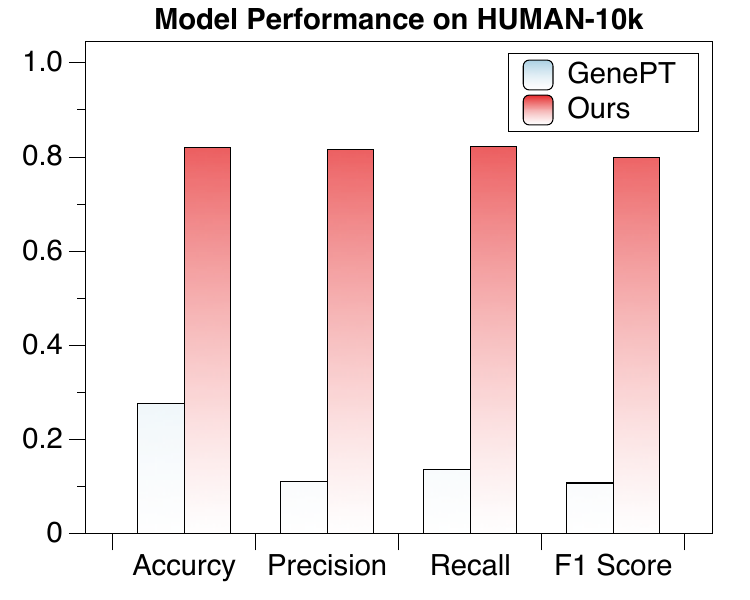}}
\subfigure[MOUSE-13k]{
    \includegraphics[width=0.45\linewidth]{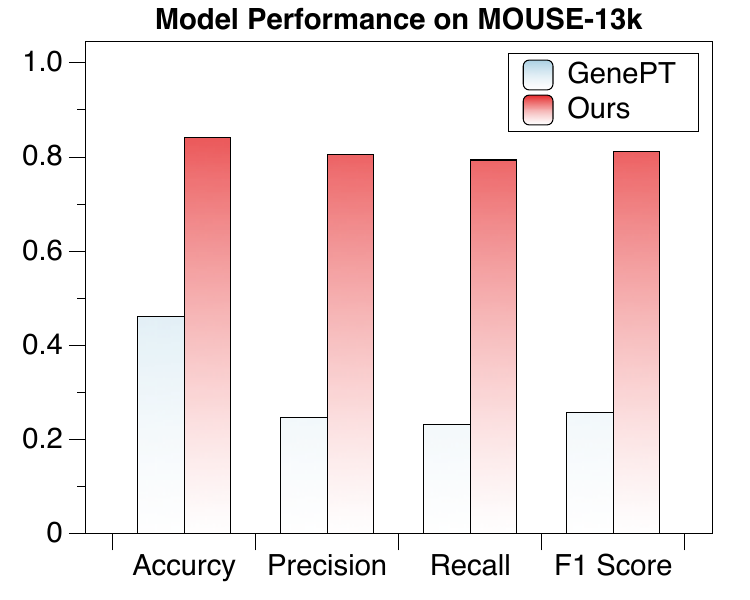}}
\caption{The performance comparison between \model\  and GenePT}
\vspace{-0.5cm}
\label{fig:1}
\end{figure}

Figure~\ref{fig:1} reported the classification performance of GenePT and \model\ on the HUMAN-10k and MOUSE-13k. 
We used four classification metrics, accuracy, precision, recall, and F1 score to evaluate two methods.
We can observe that \model\ outperformed GenePT on two datasets with a huge margin.
According to the fact that the two compared methods have the same initial gene embedding, we speculate that the common knowledge from the large language model could provide a better-supervised signal for the downstream task training, thus resulting in a better performance. 

\begin{figure}
\centering
\includegraphics[width=0.48\textwidth]{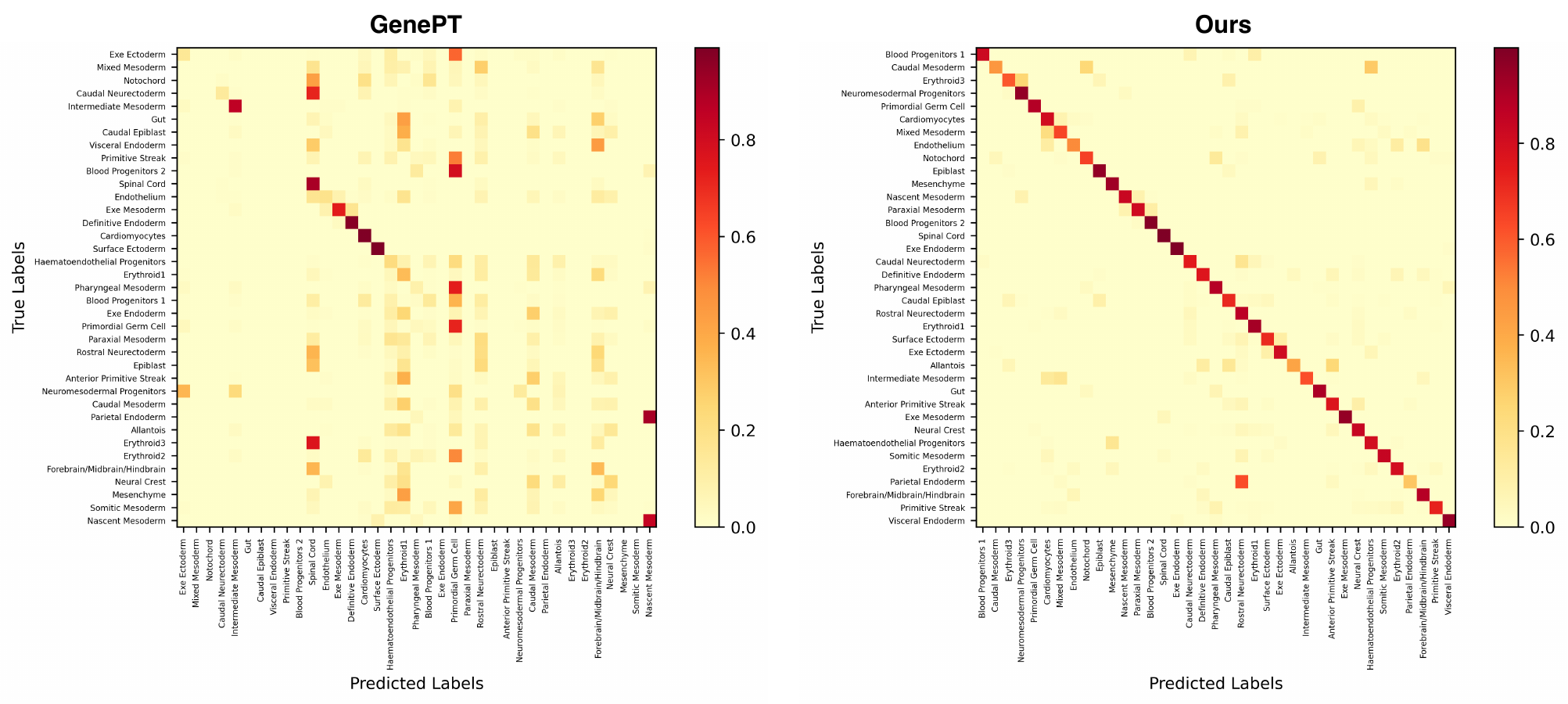}
\caption{The confusion matrix of each method on MOUSE-13k.}
\vspace{-0.3cm}
\label{fig:2}
\end{figure}


\begin{figure}
\centering
\includegraphics[width=0.45\textwidth]{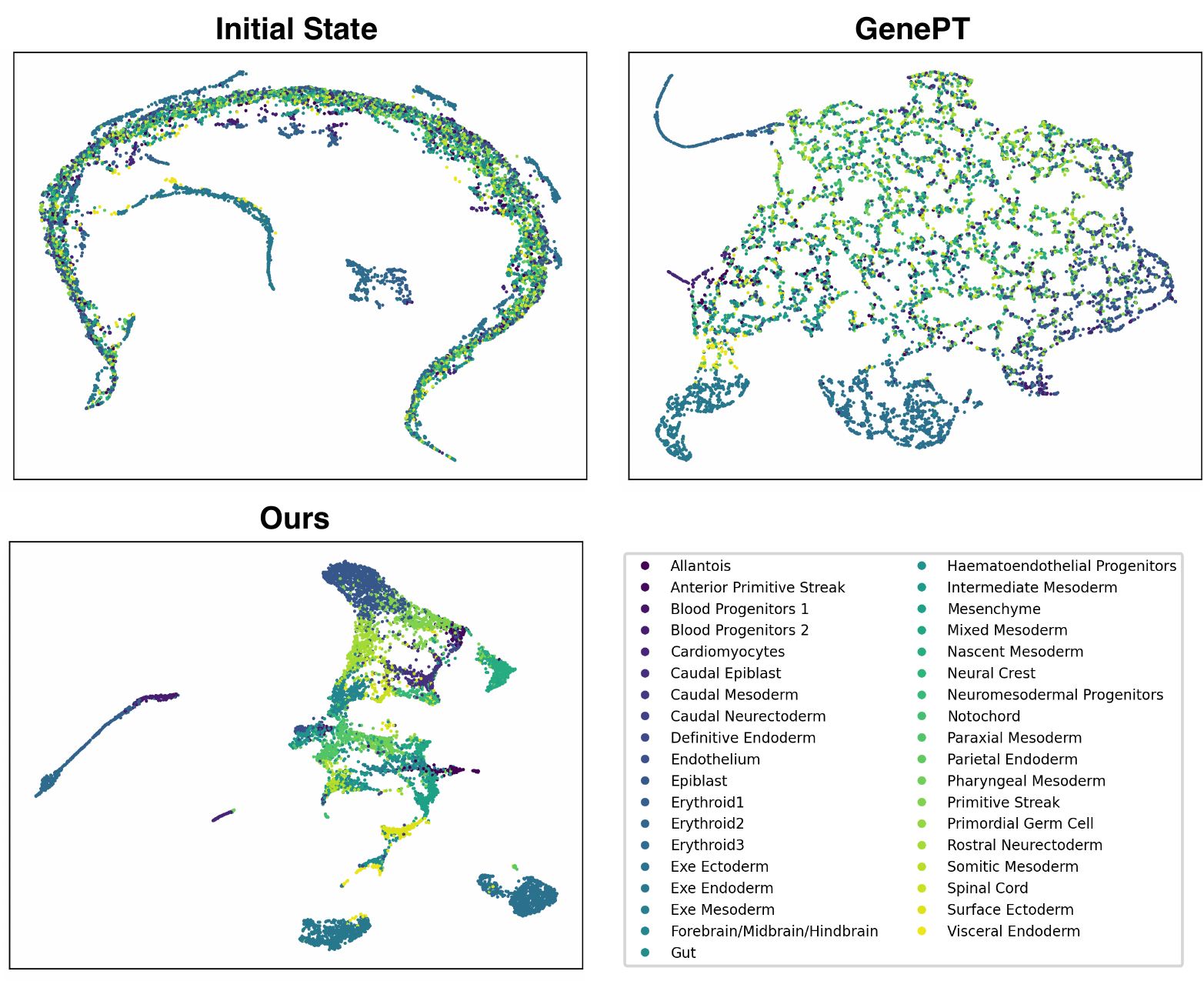}
\caption{The UMAP illustration of the cell representation from initialization, GenePT, and \model\ on MOUSE-13k.}
\vspace{-0.5cm}
\label{fig:3}
\end{figure}
To step further, we reported the confusion matrix of each method on dataset MOUSE-13k, to show the difference between the prediction result and the true label. 
From the left part of Figure~\ref{fig:2}, we could observe that most prediction results were scattered across the matrix, indicating a high degree of misclassification. 
The lighter regions of the diagonal suggest that the model frequently confuses certain cell types with others, highlighting deficiencies in either feature representation or classification capability. 
In contrast, the right part of the matrix presents a starkly different scenario, where the majority of predictions align closely with the diagonal, thus reflecting a high concordance between predicted and actual labels. 
This phenomenon denotes a significant improvement in classification accuracy, attributable to the sophisticated feature encoding and contextual understanding afforded by the prompt-based training method employed with the large language model.

\subsection{Study of the Cell Embedding Visualization}
Figure~\ref{fig:3} reported the UMAP visualization of the cell embedding from the initial state, GenePT, and \model. In the initial state of the MOUSE-13k datasets, as observed on the left side of the visualization, each cell types cluster together, exhibiting poor separability. 
The middle visualization, i.e. GenePT, shows some improvement in terms of separability among different categories. However, several cell types remain interspersed.
In contrast, the right visualization demonstrates a markedly superior clustering effect. 
Cells of the same type exhibit exceptional aggregative properties, suggesting a high degree of intra-class similarity and inter-class divergence. 
The underlying driver is that our proposed method leverages the broad, generalized knowledge inherent in LLMs to provide more effective supervisory signals, thereby enhancing the separability of learned cell embeddings.

\section{Conclusion}
This study represents a preliminary stride in the application of LLMs for enhancing gene and cell-level representations. 
Our proposed \model\ harnesses the expansive knowledge base and sophisticated understanding inherent in LLMs to interpret and categorize cell types in gene expression data.
The superior performance of \model\ over GenePT underscores the efficacy of integrating common knowledge from LLMs into the domain of gene expression analysis. 

\begin{acknowledgement}
This work is partially supported by the Postdoctoral Fellowship Program of CPSF (No.GZC20232736), the China Postdoctoral Science Foundation Funded Project (No.2023M743565), the Young Elite Scientists Sponsorship Program by BAST, and the Special Research Assistant Funded Project of the Chinese Academy of Sciences.
Cong Li and Meng Xiao contributed equally to this work.
\end{acknowledgement}
\bibliographystyle{fcs}
\bibliography{ref}

\end{document}